\documentclass[journal=jctcce,manuscript=article]{achemso}

\newcommand{\onlinecite}[1]{\hspace{-1 ex} \nocite{#1}\citenum{#1}} 
\DeclareUnicodeCharacter{308}{{\"u}}
\usepackage{graphics,graphicx,amsfonts,amsmath,amsbsy,amssymb,color}
\usepackage{verbatim}  
\usepackage{psfrag}  
\usepackage{tabularx}
\usepackage{xmpmulti}
\usepackage{hyperref}
\usepackage{marginnote}
\usepackage{subfigure}
\usepackage{paracol}
\usepackage{xparse}
\usepackage{layouts}
\usepackage{afterpage}
\usepackage{braket}
\usepackage{paralist}
\usepackage{bm}
\usepackage{url}
\usepackage[normalem]{ulem}
\usepackage{color}
\usepackage[colorinlistoftodos]{todonotes}
\usepackage[american]{babel} 

\usepackage[table]{colortbl}%
\definecolor{Tina}{RGB}{32,207,57}
\newcommand{\refeq}[1]{{Eq.~(\ref{#1})}}
\newcommand{\reffig}[1]{{Fig.~\ref{#1}}}
\newcommand{\refsec}[1]{{Sec.~\ref{#1}}}
\newcommand{\reftab}[1]{{Table~\ref{#1}}}

\newcommand{\fcut}{\ensuremath{f_\mathrm{cut}}}
\newcommand{\fcutE}{\ensuremath{f_\mathrm{cut}^{(E)}}}
\newcommand{\fcutM}{\ensuremath{f_\mathrm{cut}^{(M)}}}

\newcommand{\Ia}{Scheme 1A}
\newcommand{\Ib}{Scheme 1B}
\newcommand{\IIa}{Scheme 2A}
\newcommand{\IIb}{Scheme 2B}
\newcommand{\III}{Scheme 3}

\newcommand{\N}[1]{\ensuremath{N^{{#1}}}}

\title{How the exchange energy can affect the power laws used to extrapolate the coupled cluster correlation energy to the thermodynamic limit}%

\author{Tina~N.~Mihm}
\affiliation{Department of Chemistry, University of Iowa}

\author{Laura~Weiler}%
\affiliation{Department of Chemistry, University of Iowa}

\author{James~J.~Shepherd}%
\affiliation{Department of Chemistry, University of Iowa}%
\email{james-shepherd@uiowa.edu}
\date{\today}

\hypersetup{hyperindex,breaklinks,colorlinks=true,linkcolor=blue,citecolor=blue,urlcolor=blue,filecolor=blue}

\begin{document}

\begin{abstract}

Finite size error is commonly removed from coupled cluster theory calculations by $N^{-1}$ extrapolations over correlation energy calculations of different system sizes ($N$), where the $N^{-1}$ scaling comes from the total energy rather than the correlation energy. 
However, previous studies in the quantum Monte Carlo community %
suggest an exchange-energy-like power law of $N^{-2/3}$ should also be present in the correlation energy when using the conventional Coulomb interaction.
The rationale for this is that the total energy goes as $N^{-1}$ and the exchange energy as $N^{-2/3}$; thus, the correlation energy should be a combination of these two power laws.
Further, in coupled cluster theory, these power laws are related to the low $G$ scaling of the transition structure factor, $S(G)$, which is a property of the coupled cluster wavefunction calculated from the amplitudes.
We show here that data from coupled cluster doubles calculations on the uniform electron gas fit a function with a low $G$ behavior of $S(G)\sim G$. The pre-factor for this linear term is derived from the exchange energy to be consistent with an $N^{-2/3}$ power law at large $N$. 
Incorporating the exchange structure factor into the transition structure factor results in a combined structure factor of $S(G)\sim G^2$, consistent with an $N^{-1}$ scaling of the exchange-correlation energy.
We then look for the presence of an $N^{-2/3}$ power law in the energy. To do so, we first develop a plane-wave cutoff scheme with less noise than the traditional basis set used for the uniform electron gas.
Then, we collect data from a wide range of electron numbers and densities to systematically test five methods using $N^{-1}$ scaling, $N^{-2/3}$ scaling, or combinations of both scaling behaviors.
We find that power laws that incorporate both $N^{-1}$ and $N^{-2/3}$ scaling perform better than either alone, especially when the pre-factor for $N^{-2/3}$ scaling can be found from exchange energy calculations.

\end{abstract}

\maketitle

\section{Introduction}

{%

There has been a recent push towards developing wave-function-based methods such as coupled cluster theory for solids.~\cite{manby_extension_2006, stoll_incremental_2009, voloshina_correlation_2007, booth_towards_2013, gruneis_natural_2011, irmler_duality_2019, gruber_applying_2018, irmler_particle-particle_2019, hummel_low_2017, zhang_coupled_2019, gruneis_efficient_2015,lewis_ab_2019,mcclain_gaussian-based_2017, motta_hamiltonian_2019, pulkin_first-principles_2020, sun_gaussian_2017, mihm_shortcut_2021, neufeld_ground-state_2022, shepherd_many-body_2013, wang_excitons_2020, callahan_dynamical_2021} 
A long-term goal of this work is to provide highly accurate energy calculations for materials design. 
Coupled cluster has been growing in popularity for solid state calculations due to its ability to obtain the correlation energy (i.e., $E_\mathrm{total} - E_\mathrm{HF}$, where $E_\mathrm{HF}$ is the Hartree--Fock energy) in a versatile and systemically-improvable way.
However, one of the main issues facing coupled cluster is that the energies show slow, polynomial scaling as a function of both system size, $N$, and k-points when converging to the thermodynamic limit (TDL)---the limit of an infinite atom or particle number. %
As most energy calculations gain meaningful insight about the system at the thermodynamic limit, it is imperative that we know the exact rate at which the coupled cluster correlation energies approach this limit. %

Recent advances in coupled cluster theory have made coupled cluster single and doubles (CCSD) calculations for solids seem increasingly routine,
~\cite{mihm_shortcut_2021, neufeld_ground-state_2022, shepherd_many-body_2013, wang_excitons_2020, callahan_dynamical_2021} 
overcoming numerical convergence issues with small denominators, the divergences of perturbative methods, and technological barriers. %
In our previous studies, we have found that CCSD is a reliable way to study finite size effects (FSE) for the coupled cluster hierarchy of methods, especially when basis set errors can be effectively controlled.~\cite{mihm_accelerating_2021, mihm_power_2021, mihm_optimized_2019, mihm_shortcut_2021, shepherd_communication:_2016, weiler_machine_2022} %
In turn, the study of finite size effects is important in ensuring that coupled cluster theory is generally useful for energy calculations of solids.

A popular way to address the cost scaling issue and obtain TDL energy estimates from coupled cluster calculations for smaller system sizes is to perform an extrapolation to the TDL. 
The TDL-extrapolated energy is typically calculated by running increasingly large system sizes, and then fitting the energies at the larger $N$ to the function: $E_N = \lim_{N\rightarrow\infty}(E_{TDL}+mN^{-\gamma})$. 
Here, $N$ is the system size and refers to a number of electrons.
The number of $k$-points, $N_k$, can also be used. 
The variable $\gamma$ defines the convergence rate.
If $\gamma$ is known exactly for large $N$, the TDL energy can be estimated more accurately.   
For other energies, such as the correlation energy, there is an ongoing discussion in the literature as to the exact value of both $\gamma$ and the form of the extrapolation equation itself. 

A commonly-assumed convergence rate for the correlation energy is $N^{-1}$, ~\cite{shepherd_many-body_2013, liao_communication:_2016, gruber_applying_2018}
the same as the total energy relationship. %
}
The \N{-1} convergence of the correlation energy, which is physically attributed to long-range van der Waals forces~\cite{marsman_second-order_2009, gruber_applying_2018, fraser_finite-size_1996}, and can be derived in the UEG \cite{drummond_finite-size_2008}, 
has been related to the low momentum limit ($G \rightarrow 0$) of the transition structure factor $S(G)$. The transition structure factor is calculated from the amplitudes of the CCSD wavefunction, and the sum over its pointwise product with the Coulomb operator in k-space yields the correlation energy. As such, its scaling at low momenta relates to the power law of the TDL extrapolation: a convergence of $S(G) \sim G^2$ predicts a power law of \N{-1}.~\cite{liao_communication:_2016, mihm_power_2021, holzmann_theory_2016, holzmann_finite-size_2011, mattuck_guide_1992} 
This also matches the ground state structure factor convergence as $G\rightarrow 0$, which has been extensively explored in the QMC and DFT literature.~\cite{chiesa_finite-size_2006,holzmann_theory_2016, gori-giorgi_analytic_2000, ortiz_correlation_1994} 

There is also very strong evidence from the QMC literature that there is a term of $\N{-2/3}$ in the correlation energy.~\cite{ruggeri_correlation_2018, drummond_finite-size_2008, azadi_correlation_2022} %
The \N{-2/3} scaling first appears in the exchange energy convergence into the TDL when using an Ewald interaction.~\cite{fraser_finite-size_1996}
To reach an \N{-1} scaling in the total energy, it is reasonable to assume that the correlation energy must have an equal and opposite term in \N{-2/3}.
For periodic coupled cluster theory, an \N{-2/3} convergence in the correlation energy would mean that there is a $S(G) \sim G$ scaling behavior in the transition structure factor that has not yet been identified. 
This leaves open the question as to whether the CCSD energy has analogous relationships to energies from QMC.

In this study, we will identify how an \N{-2/3} term arises in the CCSD correlation energy by first analyzing the transition structure factor. 
We show that the correlation-only transition structure factor (from finite CCSD calculations) fits a functional form with $S(G)\sim G$ scaling in the limit as $G\rightarrow 0$. 
We then show that the term in $G$ can be cancelled by the exchange component of the ground state structure factor, giving rise to the expected overall $S(G) \sim G^2$ scaling of the ground state structure factor. 
We present numerical and analytical results to show how this scaling in $S(G)$ affects energy extrapolations to the thermodynamic limit, paying particular attention to comparing \N{-1} and \N{-2/3} extrapolations in practical contexts. 
We argue in favor of incorporation of an \N{-2/3} term in the correlation energy extrapolation provided that its prefactor can be derived from exchange-energy calculations.

\section{Methods}

\subsection{Coupled cluster theory and the uniform electron gas}

All calculations in this paper were performed using coupled cluster theory, where we followed the methods detailed in our previous papers.~\cite{shepherd_range-separated_2014, shepherd_coupled_2014,shepherd_many-body_2013} Here, we will just outline some of the main methodological details for clarity. In coupled cluster theory, an exponential ansatz is used for the wavefunction: $\Psi = e^{\hat{T}}\Phi_0$, where $\Phi_0$ is the ground state wavefunction, typically taken to be the Hartree--Fock wavefunction, and $\hat{T}$ is the excitation operator. This wavefunction is then used to find the coupled cluster correlation energy by projection, i.e., $E = \langle \Phi_0 |H| e^{\hat{T}}\Phi_0 \rangle$.
As the work presented here is performed in the uniform electron gas (UEG) where singles excitations are zero due to conservation of momentum, 
we typically truncate the $T$-amplitudes to just the doubles to give the coupled cluster doubles (CCD) energy. This energy, then, is calculated using the following equation: 
\begin{equation}
E_\mathrm{c} = \frac{1}{4} \sum_{ijab} t_{ijab}\bar{v}_{ijab}
\label{correlation_energy}
\end{equation}
where $t_{ijab}$ are the T-amplitudes only for the doubles excitations, and $\bar{v}_{ijab}$ %
are the antisymmetrized four-index electron repulsion integrals. Per convention, $i$ and $j$ index occupied orbitals and $a$ and $b$ index virtual orbitals for a finite basis set.   
Following a similar derivation to the one in the work by Liao and Gr{\"u}neis~\cite{liao_communication:_2016}, this energy expression is equivalent to one rewritten in terms of the transition structure factor, $S({\bf G})$: 
\begin{equation}
E_\mathrm{c} = \frac{1}{2} \sum^\prime_{{\bf G}} S({\bf G}){v}({\bf G}) .
\label{correlation_energy_with_Sg}
\end{equation}
The $^\prime$ symbol denotes that the sum does not include the $G=0$ term. 
The structure factor is given by: $S({\bf G}) = \sum_{ijab} (2T_{ijab}–T_{jiab}) \Theta_{ijab}({\bf G})$. 
The $T$ appears in place of $t$ to reflect that the indices are now spatial orbitals. 
The $\Theta_{ijab}({\bf G})$ %
indicates that it only goes over the excitations that are related to the momentum transfers, ${\bf G}$, with ${G}$ being the magnitude of the momentum transfer between the {$i,j$} to {$a,b$} excitation (i.e., $|{\bf G}| = G$).
The additional factor of $1/2$ comes from the convention we used for the UEG structure factors (for consistency with our prior work).~\cite{mihm_power_2021} %

Our electron gas also follows the same set-up as described in our previous work,~\cite{shepherd_coupled_2014, mihm_power_2021}
with the exception that this work also contains open-shell electron configurations. 
For our UEG system, we use a simple three-dimensional cubic box with electron numbers that that correspond to open- and closed-shell configurations (relative to a grid centered at the $\Gamma$-point). 
The volume of the box is calculated using the Wigner–Seitz radius, $r_s$, such that $\Omega = L^3 \approx \frac{4}{3}\pi r_s^3 N$, where $L$ is the length of one side of the box 
We work exclusively in a plane wave basis set for our UEG calculations, where all the orbitals are described using the relationship $\phi_j \propto \exp({\sqrt{-1} {\bf k}_j \cdot {\bf r}})$, where ${\bf k}_j$ is a momentum vector for orbital $j$, and ${\bf r}$ is the electron coordinate.  
Ewald interactions are employed for the periodic boundary condition calculations, as per convention, which causes $1/{G}^2$-type matrix elements to appear in the electron repulsion integrals, $v_{ijab}$.
As there is conservation of momentum in the UEG, only the excitations that correspond to ${\bf G}$’s that meet the requirement ${\bf k}_a-{\bf k}_i={\bf k}_j-{\bf k}_b={\bf G}$ are used. 
As with our previous work, we explicitly calculate and include the Madelung term, $v_M$.~\cite{fraser_finite-size_1996, mihm_power_2021} We also use a finite basis set defined by the $M$ orbitals that lie inside a kinetic energy cutoff $E_{cut, m} = \frac{1}{2}k_\mathrm{cut}^2$.
The Hartree--Fock eigenvalues for the occupied and virtual orbitals follow the same conventions as our previous work~\cite{shepherd_coupled_2014, mihm_power_2021}, and are lowered in energy by the $v_M$ term. In the thermodynamic limit, $v_M\rightarrow 0$.

\subsection{Connectivity twist averaging}

Twist averaging is typically used to help reduce finite size effects by reducing the fluctuations in the wavefunction as the system converges to the thermodynamic limit (TDL).~\cite{ lin_twist-averaged_2001, gruber_ab_2018, gruber_applying_2018, drummond_finite-size_2008, liao_communication:_2016, maschio_fast_2007, zong_spin_2002, pierleoni_coupled_2004, mostaani_quantum_2015, azadi_efficient_2019}
This is typically accomplished by applying a series of offsets to the orbitals called twist angles, ${\bf k}_s$, 
{such that $\phi_j \propto \exp(\sqrt{-1}\, ({\bf k}_j - {\bf k}_s)\cdot {\bf r})$, and then averaging the correlation energy over each twist angle:
\begin{equation}
\langle E_\mathrm{corr} \rangle_{{\bf k}_s}=\frac{1}{N_s}\sum_{t=1}^{N_s}E_\mathrm{corr}({\bf k}_{s,t})
\end{equation}
Here, the average involves a sum over $N_s$ coupled cluster calculations (where $N_s$ is the number of twist angles). This increases the cost of running twist-averaged coupled cluster by a factor of $N_s$.}

In order to help reduce this cost while still obtaining twist-averaged energies for larger systems, we instead use our connectivity twist averaging (cTA) method, which was introduced in other studies.\cite{mihm_optimized_2019, mihm_accelerating_2021} 
{With this method, we find a special twist angle for each calculation that reproduces the twist-averaged energies.}
The method works through evaluating the momentum transfer vectors between the occupied and virtual space, dubbed the “connectivity”. {These momentum transfer vectors are used to find} the twist angle that most closely matches the averaged connectivity {using a residual difference calculation}. 
As each system size will have a different connectivity, a special twist angle must be selected individually for each system. 
The advantage here is that we are now running a calculation using a single twist angle for each system, but reproducing twist-averaged energies, lowering the cost of obtaining twist averaged energies by a factor of $N_s$. With this cost reduction, we can then obtain twist-averaged energies for much larger systems, which is vital to our work presented here. 

\subsection{An improved $f_\mathrm{cut}$ basis set scheme}
\label{CCD_BS_Correction}

In our previous work,~\cite{mihm_power_2021,shepherd_many-body_2013} we used a basis set scheme that employs a cutoff factor ($f_\mathrm{cut}$) to truncate the basis set to a given number of orbitals, $M$. This cutoff factor was chosen such that $f_\mathrm{cut}=E_{\mathrm{cut},M}/E_{\mathrm{cut},N}$, where $E_{\mathrm{cut},M}$ refers to the energy cutoff for the basis set with $M$ orbitals and $ E_{\mathrm{cut},N}$ refers to the energy cutoff for the system size containing $N$ electrons. With this method, the $E_{\mathrm{cut},M}$ was calculated manually before being provided to our coupled cluster code for use in truncating the basis set. This basis set scheme will be referred to as $\fcutE$ in the text, where the $(E)$ is referencing the use of the energy cutoffs to truncate the basis set. 

A more precise way to truncate the basis set is to control the number of basis functions per electron with the benefit of allowing the automatic adjustment of the basis set when the electron number changes. 
In this new basis set scheme, 
we use our chosen $f_\mathrm{cut}$ and the number of electrons to calculate $M$ on the fly using the following equation: 
\begin{equation}
M = (\fcutM)^{3/2} N.
\label{fcut_prime_eq}
\end{equation}
Here, $\fcutM$ is the ratio of basis functions per electron re-scaled into energy units (using the $3/2$ power). With this new method of truncating the basis set, we get a more accurate number of orbitals given our target $f_\mathrm{cut}$. One of advantages of this new $\fcutM$ basis set scheme is that, unlike with our previous $\fcut$ scheme, we are not limited to only the closed-shell system sizes---system sizes determined by symmetry such as $N=14, 38, 54, etc$---as the basis set cutoff is now calculated directly. This allows us to use open-shell systems that break symmetry such as $N = 26, 46, 60, etc$. %
\begin{figure}

\includegraphics[width=0.49\textwidth,height=\textheight,keepaspectratio]{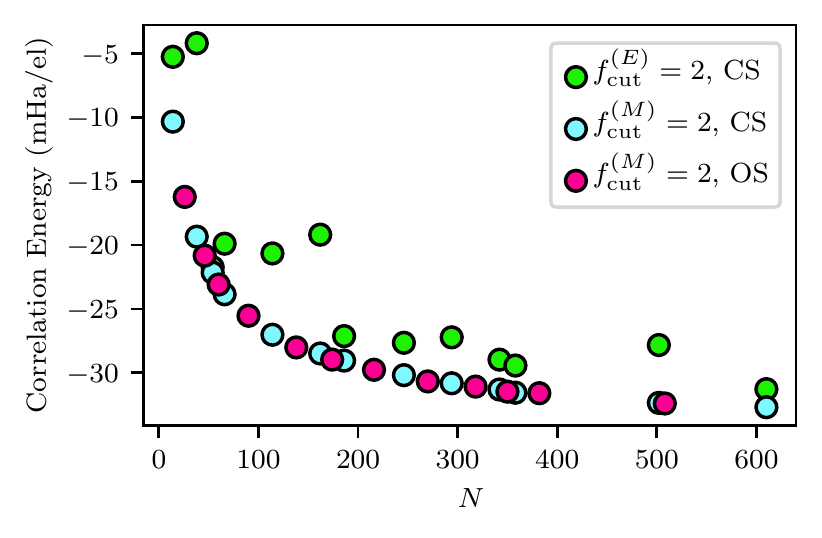}

\caption{Comparison between the correlation energies for the $\fcutE = 2$ and $\fcutM = 2$ basis sets {are} shown for a range of system sizes at a density of $r_s = 1.0$. 
The $\fcutE = 2$ basis set energies are shown for only closed-shell (CS) system sizes, while the $\fcutM = 2$ basis set energies are shown for both closed-shell and open-shell (OS) systems sizes.  
This comparison shows that both the $\fcutM = 2$ basis sets have less noise in their convergence compared to the $\fcutE = 2$ basis set, resulting in
a smoother convergence to the TDL. 
}
\label{fig:BasisSet_comp_fig}
\end{figure}

Figure \ref{fig:BasisSet_comp_fig} shows the results of a comparison between the energies for the $\fcutE$ and $\fcutM$ basis sets at an $r_s = 1.0$. 
For both $\fcutE$ and $\fcutM$, a cutoff factor of $2$ was used for a range of electron numbers from $N = 14$ to $N = 730$ for the closed-shell systems and from $N = 26$ to $N = 508$ for the open-shell systems.
In \reffig{fig:BasisSet_comp_fig}, the energies for the two basis set schemes are shown graphed against increasing electron number. As can be seen in the figure, the energies for both the closed-shell and the open-shell systems using the $\fcutM$ basis set show a smoother convergence to the TDL than the energies for the closed-shell systems that used the $\fcutE$ basis set. These results support the idea that the $\fcutM$ basis set helps reduce basis set incompleteness error (BSIE) that causes changes in the finite size error when the electron number changes. Overall, \fcutM provides a smoother TDL convergence.

\subsection{Correcting basis set incompleteness error}

Basis set incompleteness error (BSIE) was handled in the normal way, 
\cite{mihm_power_2021} through deriving a correction to the BSIE from the complete basis set (CBS) limit. This helps ensure that the energies are converged with respect to the basis set before they are extrapolated to the TDL. 
Since the extrapolations in $M$ and $N$ tend to be independent and commute~ \cite{shepherd_communication:_2016}, we can calculate a basis set correction by choosing a fairly large electron number (here $N = 216$) and running calculations with increasing basis set sizes. These energies are then extrapolated to the CBS limit, and a basis set correction is calculated using the following equation: 
\begin{equation}
\Delta E_\mathrm{CBS} = E_\mathrm{CBS} - E(216, M)
\label{CBS_correction}
\end{equation}
Where $M$ is our chosen basis set size determined by the $\fcutM$ (here, $\fcutM = 2$) and $E_\mathrm{CBS}$ is the energy at the CBS limit. This correction term, $\Delta E_\mathrm{CBS}$, is uniformly added to the energies for all $N$. This process is then repeated for all densities.

\subsection{Background literature on the ground-state structure factor}

In our previous work\cite{mihm_power_2021}, we fit the CCSD transition structure factor to the following function (inspired by screened MP2) which had a limiting form of $S(G) \sim G^2$ as $G\rightarrow 0$:
\begin{equation}
S_G \propto \frac{1}{(G^2+\lambda^2)^4}G^2.
\end{equation}
This equation leads to an $N^{-1}$ form for the energy as it approaches the TDL due to the $G^2$ asymptotic behavior at small $G$. 
Here, as we are investigating $N^{-2/3}$ and $N^{-1}$, we require a different functional form that incorporates $S(G) \sim G$ as $G\rightarrow 0$. %

We take as our inspiration an accurate and well-fitting functional form for the ground-state structure factor, which was proposed by Gori-Giorgi \emph{et al}~\cite{gori-giorgi_analytic_2000}.
Their functional form incorporated analytical results from the Hartree--Fock approximation (for exchange), the random phase approximation (for the low momentum region), and Quantum Monte Carlo calculations. 
The ground-state structure factor includes components that correspond to both exchange and correlation structure factors. 
For our analysis of transition structure factors, we will be using a modified form of this function that only includes the terms coming from Gori-Giorgi {\emph{et al.}}’s correlation structure factor. 

Specifically, when we fit our correlation transition structure factor, we used:
\begin{equation}
\begin{split}
    S_c(G) = e^{-B_1 G} &\bigg( -\frac{3}{4 q_F}G + C_2 G^2 + C_3 G^3 \\ 
    &+ C_4 G^4 + C_5 G^5 + C_6 G^6 + C_7 G^7\bigg) %
\end{split}
\label{Corr_SF_PGG}
\end{equation}%
This differs from the structure factor proposed by Gori-Giorgi \emph{et al}~\cite{gori-giorgi_analytic_2000} in that it does not make an attempt to separate the different spin components of the structure factor; instead, we group all of these terms together to simplify the fitting analysis. 
While those authors also fixed the number of coefficients for each $r_s$ value, we found it necessary to include $C_6$ and $C_7$ only for $r_s=1.0$. These were removed for $r_s=5.0$.  Additionally, the inclusion of $C_7$ was our own addition. 
We also neglected any treatment of the cusp condition---mainly because the cusp was not our focus in this study and we used small basis sets for these fits. 

In common with Gori-Giorgi \emph{et al}~\cite{gori-giorgi_analytic_2000}, the term that is linear in $G$ is constrained such that in the limit of low $G$, $S_c(G)\sim -\frac{3}{4q_F}G$. 
That this is constrained \emph{a priori} forces the condition that the low-$G$ limit of the exchange-correlation transition structure factor goes as $\sim G^2$. 
Unlike those authors, we did not fix any further higher-order terms, instead relying upon fitting the function to our data to determine the superlinear coefficients. 

We also used an exchange structure factor of the form: 
\begin{equation}
    S_x(G) = \left\{
    \begin{array}{ll}
    -1 + \frac{3}{4 q_F} G - \frac{1}{16 q_F^3} G^3 , & G \leq 2q_F \\
    0 , & G > 2q_F
    \end{array}
    \right. 
\label{Ex_SF_PGG}
\end{equation}
where $q_F$ is the Fermi wave vector equal to $q_F= \alpha/r_s$, where $\alpha = (9 \pi/4)^{1/3}$. 
Here, there are no fit parameters. 
The momentum transfer $G=2q_F$ is the largest momentum transfer that fits inside the Fermi sphere; thus, we are guaranteed to not have any contributions from $G$ larger than this, allowing $S_x(G) = 0$ for $G > 2q_F$. 

To make a connection between HF exchange structure factor and transition structure factor, we can note that an exchange transition structure factor can be defined: 
\begin{equation}
\begin{split}
E_x&=\frac{1}{2} \sum_{ij \in \mathrm{occ}} \frac{1}{\Omega} \frac{4\pi}{|{\bf k}_i-{\bf k}_j|^2}  \\ 
&=\frac{1}{2} \sum_G {}^\prime S_\mathrm{x}(G) V(G)
\end{split}
\end{equation} 
In this equation, the sum over $i$ and $j$ runs over occupied orbitals, and $G$ is the momentum difference between $k_i$ and $k_j$ (i.e., the magnitude of ${\bf G}= {\bf k}_i-{\bf k}_a$). %
The $^\prime$ symbol denotes that the sum does not include the $G=0$ term. 
The factor of $\frac{1}{2}$ maintains consistency with derivations from other authors, and is included due to double counting in sums over electron pairs.

\begin{figure*}

\subfigure[\mbox{}]{%
\includegraphics[width=0.45\textwidth,height=\textheight,keepaspectratio]{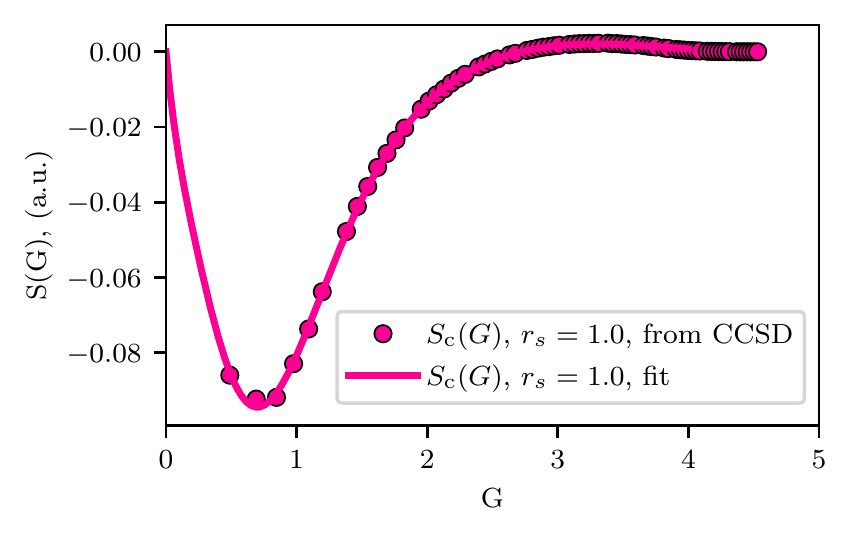}
\label{subfig:Rs1_ScF}
}
\subfigure[\mbox{}]{%
\includegraphics[width=0.45\textwidth,height=\textheight,keepaspectratio]{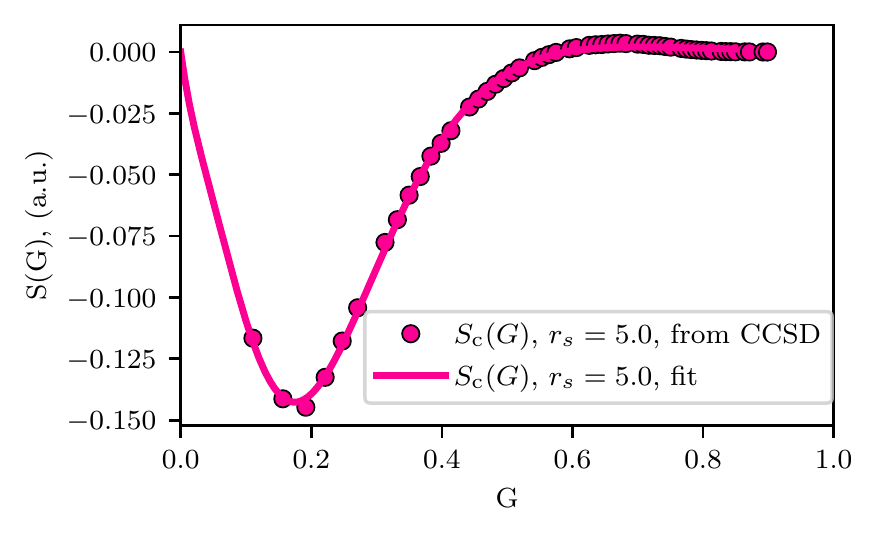}
\label{subfig:Rs5_ScF}
}

\subfigure[\mbox{}]{%
\includegraphics[width=0.45\textwidth,height=\textheight,keepaspectratio]{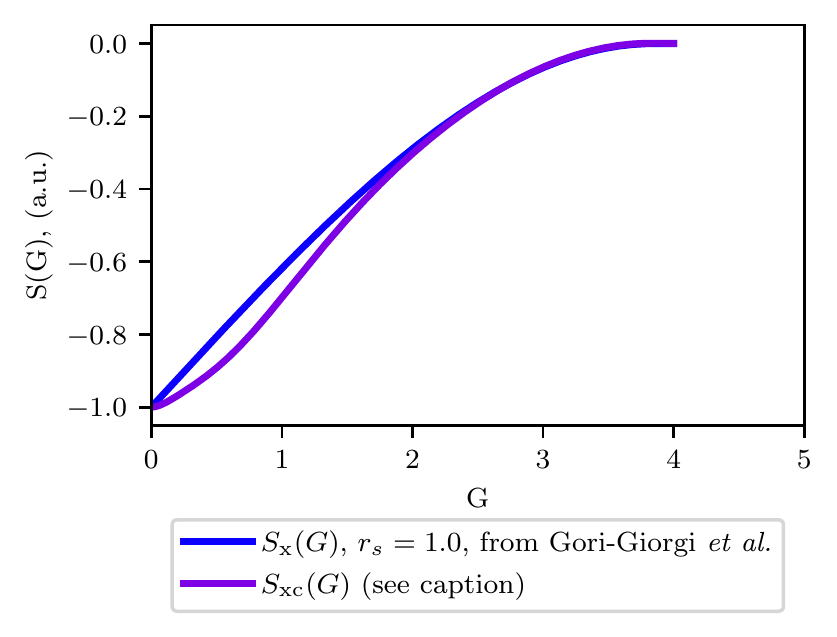}
\label{subfig:Rs1_SF}
}
\subfigure[\mbox{}]{%
\includegraphics[width=0.45\textwidth,height=\textheight,keepaspectratio]{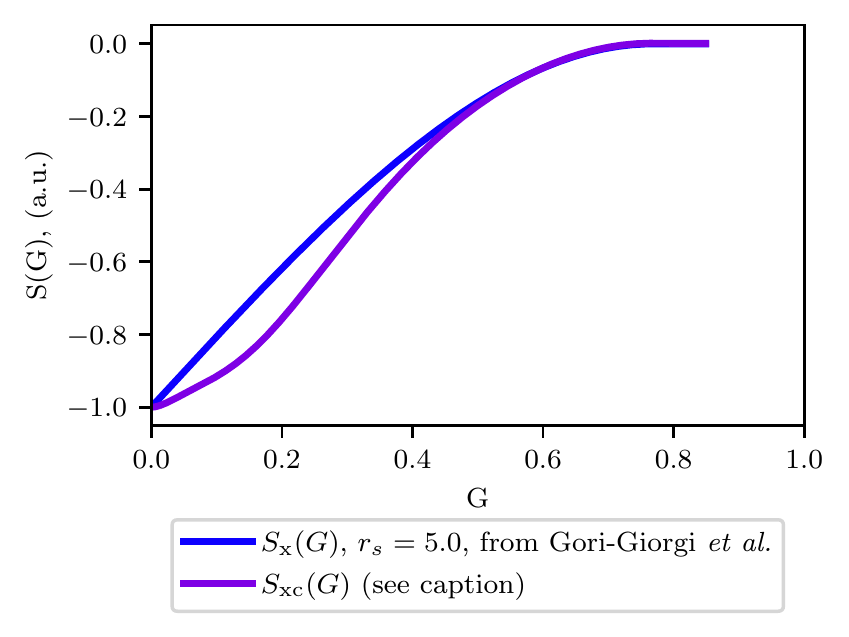}
\label{subfig:Rs5_SF}
}

\caption{
The correlation ($S_\mathrm{c}(G)$), exchange ($S_\mathrm{x}(G)$), and exchange-correlation ($S_\mathrm{xc}(G)$) transition structure factors are shown for a density of (a, c) $r_s = 1.0$ ($N=508$) and (b, d) $r_s = 5.0$ ($N=350$).
$S_\mathrm{c}(G)$ points come from a CCSD calculation, which are subsequently fit using \refeq{Corr_SF_PGG} to make a continuous function. This is then added to $S_\mathrm{x}(G)$ (defined by \refeq{Ex_SF_PGG}) to make an $S_\mathrm{xc}(G)$ line. %
The functions $S_c(G)$ and $S_x(G)$ are linear into the origin, while $S_{xc}(G)$ is quadratic. 
}
\label{fig:Paola_SF_fits}
\end{figure*}

\section{Results}

\subsection{Calculation details}
\label{Calculation_Details}
For the rest of this work, we will be working with data 
collected over a range of seven $r_s$ values. All $r_s$ values %
were run at an $\fcutM = 2$ for a range of $N$ to obtain convergence to the TDL. 

The calculations were performed on the following open shell electron numbers:
$N = 26$, 46, 60, 90, 138, 174, 216, 270, 318, 350, 382, 508, 646, 754, and 826.
Within this set, we used an electron range of $N = 26$ to $826$ for $r_s = 0.1$, $N = 26$ to $508$ for $r_s = 1.0$ and $2.0$, $N = 26$ to $350$ for $r_s = 5.0$, $N = 26$ to $270$ for $r_s = 10.0$ and $20.0$, and $N = 26$ to $216$ for $r_s = 50.0$.
The upper limit on $N$ was determined by how well the calculations converged.

For basis set corrections, we used $N=216$ and a range of basis sets from $M = 302$ to $3788$ for $r_s = 1.0, 2.0$ and $5.0$. For $r_s = 0.1$, the basis set range used was up to $M = 5590$ and for $r_s = 10.0$ and $20.0$ the basis set range went up to $M = 4548$. %

All calculations were performed using a locally-modified version of a github repository used in our previous work: http://github.com/jamesjshepherd/uegccd ~\cite{shepherd_range-separated_2014,shepherd_coupled_2014}. Hartree atomic units are used throughout.

All graphs were plotted using {\tt{matplotlib}} with {\tt{Python 3.7.3}}. 
For the extrapolations to the TDL the {\tt{numpy}} and {\tt{scipy}} libraries were used with {\tt{Python 3.7.3}}.

All fits for the extrapolation schemes were performed using the {\tt{curve\_fit}} function from the {\tt{scipy}}  library in {\tt{Python}}. The error in each TDL estimate was calculated from the variance in the fitting parameters.

\subsection{Fitting the transition structure factor and accounting for exchange}
\label{Theory_SF_analysis}
In \reffig{fig:Paola_SF_fits}, we show the transition structure factors for $r_s = 1.0$ ($N=508$) and another at $r_s = 5.0$ ($N=350$).
These show calculations of the transition structure factor from CCSD calculations using the relationships described above in \refeq{correlation_energy} and \refeq{correlation_energy_with_Sg}. %

In both \reffig{subfig:Rs1_ScF} and \ref{subfig:Rs5_ScF}, %
the raw correlation structure factor ($S_c(G)$) data is shown with our transition structure factor fit for both $r_s = 1.0$ and $5.0$ respectively. The $S_c(G)$ fit is based on a modified form of the of the ground-state structure factor fit proposed by Gori-Giorgi \emph{et al.}~\cite{gori-giorgi_analytic_2000}, which only includes the correlation components (see \refeq{Corr_SF_PGG}). %
Here, the $S_c(G)$ fit for both $r_s$ has a fixed linear term that is equal and opposite the known linear term from the exchange structure factor. 
This fixed linear term gives $S_c(G)$ a linear convergence to zero as $G\rightarrow 0.0$.
The close fit between the curve and the data demonstrate that the functional form is consistent with our data. 
We tested releasing the constraint on the size and sign of the linear term. 
Unconstrained fits of both results in a curve that is much less well fit, but constraining the sign of the linear term results in a reasonable fit with a coefficient of the same order of magnitude as the original linear term. %

Fitting the transition structure factor also allows us to show what happens when the exchange structure factor is included, which is shown in \reffig{subfig:Rs1_SF} and \ref{subfig:Rs5_SF}. %
Here, the exchange structure factor is plotted using \refeq{Ex_SF_PGG} %
These are then combined with the transition structure factor to make the exchange-correlation transition structure factor: 
\begin{equation}
S_\mathrm{xc}(G) = S_\mathrm{c}(G) + S_\mathrm{x}(G).
\end{equation}
Here, the $S_\mathrm{c}(G)$ is taken from fitting \refeq{Corr_SF_PGG} to CCSD data, and $S_\mathrm{x}(G)$ is from the analytical form given in \refeq{Ex_SF_PGG}.
When $S_c(G)$ and $S_x(G)$ are added together, the linear terms cancel by construction (compare \refeq{Corr_SF_PGG} with \refeq{Ex_SF_PGG}), leaving the exchange-correlation structure factor with a quadratic convergence to zero as $G\rightarrow 0.0$. 
The success of these fits goes some way to demonstrating that the functions proposed by Gori-Giorgi \emph{et al}~\cite{gori-giorgi_analytic_2000} appropriately model the low $G$ regime of the transition structure factor for the coupled cluster correlation energy.

\begin{figure}[]
\includegraphics[width=0.49\textwidth,height=\textheight,keepaspectratio]{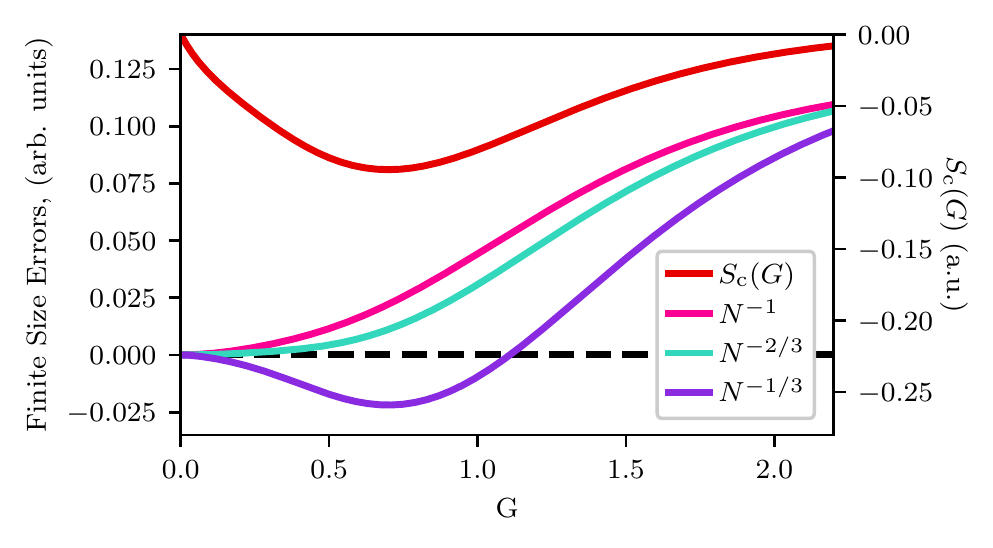}

\caption{Our analytical fit for the correlation transition structure factor for $r_s = 1.0$ is used to analyze the convergence of the finite size errors (FSE) to the thermodynamic limit using three different power laws.
\N{-1/3} shows an over estimate of the FSE as $G$ approaches $0.0$, while \N{-1} and \N{-2/3} show very similar convergence into the origin.
The units of the extrapolations are in arbitrary units due to the proportionality relationship in \refeq{SG_integration}.
}
\label{fig:FSE_Anaysis_fig}
\end{figure}

\subsection{Comparing \N{-2/3} and \N{-1} extrapolations analytically}
\label{Sg_Analysis_Extrap}

With a continuous fit for the structure factor, we are able to examine how finite size effects in the correlation energy converge as the system approaches the TDL. 
Here, we will follow the same derivation as our previous paper  \cite{mihm_power_2021} using our analytically-derived transition (correlation) structure factor to analyze the FSE for the \N{-2/3} and \N{-1} TDL convergence rates. All symbolic manipulations and fits in this section were performed in Mathematica.~\cite{wolfram_research_inc_mathematica_2019}  

Using the relationship between the energy and the transition structure factor given in \refeq{correlation_energy_with_Sg}, we start by integrating over the part of the analytical form of the correlation transition structure factor shown in \refeq{Corr_SF_PGG}, that spans from zero to the minimum $G$ present in our data, $G^\prime$, to obtain the finite-size error present in the correlation energy:
\begin{equation}
F(G^\prime) \propto \int_{0}^{G^\prime} S_c(G) v(G) \, G^2 dG .
\label{SG_integration}
\end{equation}
Here, the factor of $G^2$ comes from the $G$-space volume element in 3D. 
As we are only interested in relative errors, the expression here does not consider any constant prefactors, which are considered in the next section.

It is now possible to estimate the amount of finite size error which is left after extrapolation. 
Extrapolation consists of fitting the energies to a linear function of the system size (e.g. \N{-2/3}, equivalent to $G^2=(\frac{2\pi}{L})^2$). %
Thus, the removed FSE can be related to the derivative of the function form of the energy ($F(G^\prime)$):  
\begin{equation}
\Delta E_\mathrm{FSE} \propto G^\prime\,^2 \frac{d}{d G^\prime\,^2} F(G^\prime), 
\label{SG_FSE_slope}
\end{equation}  
and the energy left after extrapolation is: 
\begin{equation}
F(G^\prime)- \Delta E_{FSE} .
\label{SG_FSE_remaining}
\end{equation}  
Here, it is important to note that $G^\prime\,^2=(\frac{2\pi}{L})^2\propto N^{-2/3}$. 
We found analagous expressions for \N{-1/3} and \N{-1}.~\cite{mihm_power_2021}
The overall result of this analysis is to be able to test different power laws and how they fit to our structure factor, which was forced to behave as $S_\mathrm{c}\sim G$ as $G\rightarrow 0$.

Figure \ref{fig:FSE_Anaysis_fig} shows the result of this analysis for the \N{-1}, \N{-2/3} and \N{-1/3} power laws. 
The \N{-2/3} power law shows the best convergence into the TDL, as would be expected given that we fixed the behavior of the structure factor to $S_\mathrm{c}\sim G$ as $G\rightarrow 0$.
While the \N{-2/3} and \N{-1} power laws are both reasonably similar, 
the \N{-2/3} power law does show slightly improved convergence to the TDL. 
Here, we can see that the \N{-1/3} power law overshoots the TDL as the finite size error converges to zero. This is in line with the results from our previous study.~\cite{mihm_power_2021}
This is expected from our use of a structure factor model that goes linearly in $G$ as $G \rightarrow 0$. 

\subsection{Calculating the TDL energy using an interpolation}
\label{Calculating_the_Analytical_TDL}

Given that we now have an analytical form of both the correlation transition structure factor that agree with our current $S(G)$ data as shown in \reffig{fig:Paola_SF_fits}, we should be able to integrate over $S(G)$ to get an analytical TDL estimate---in other words, integrating over \refeq{SG_integration} with the constant prefactors included. This is analogous to the approach by Liao and Grueneis.~\cite{liao_communication:_2016} %

Following this approach, we can calculate the correlation energy as:
\begin{equation}
    E_\mathrm{TDL} = \frac{1}{2}\bigg(\frac{L}{2\pi}\bigg)^3 \int S_c(G) \bigg( \frac{1}{L^3} \frac{4\pi}{G^2} \bigg) 4\pi G^2 dG 
\label{Analytical_TDL}
\end{equation}
Here, the $\frac{1}{L^3} \frac{4\pi}{G^2}$ is the electron repulsion integral in reciprocal space and $4\pi G^2$ is the 3D volume element. The $(\frac{L}{2\pi})^3$ term is inverse of the k-space volume of one grid point. 
The additional factor of $1/2$ is to maintain consistency with \refeq{correlation_energy_with_Sg}.

The energy produced from this integral, $E_\mathrm{TDL}$, is in units of Ha/el. 
It is important to note here that the $E_\mathrm{TDL}$ from the above equation still contains basis set incompleteness error, so we need to apply the same uniform basis set correction that we used on the correlation energies. 
For $r_s$ of $1.0$ and $5.0$, we get TDL energies of $-56.52$ mHa/el and $-22.64$ mHa/el, respectively, after all corrections. %
With these analytical fits, we can now assess which extrapolation scheme from the following section gives the best TDL estimate for our data.
\subsection{Overview of extrapolation schemes to reach the TDL}
\label{Extrap_Schemes_TDL}

The goal of the rest of this manuscript is to use data from a range of calculations to compare different power laws for their effectiveness at converging exchange and correlation energies to the TDL. From the above analysis, our hypothesis is that \N{-2/3} is the limiting power law to the TDL, which replaces the \N{-1} power law that we and other authors have used in the recent history for extrapolating the correlation energy.
We wish to explore other questions, such as when it is best to use \N{-2/3} compared with other power laws, and whether extrapolations that have more than one power law are effective.

\label{Test_Schemes}

We will be comparing five different ways to extrapolate the correlation energy to the thermodynamic limit. 
Here we give a complete description of how each extrapolation was performed along with a label for each scheme that will be used throughout the rest of this work. 
We have five schemes in total. Each has a number and may have a letter.
The number of the scheme refers to the number of variables used in the fit, while the letter distinguishes different power laws with the same number of variables. 
For example, {\Ia} and {\Ib} both have one variable used in their fit and they differ because the limiting power law they use is different.

In {\Ia}, we use a straightforward $N^{-1}$ convergence rate to extrapolate our basis-set-corrected twist-averaged correlation energies to the TDL. This is the most common way of extrapolating the correlation energy used in the literature, though in some cases has other supporting functions.\cite{liao_communication:_2016, liao_comparative_2019, marsman_second-order_2009, gruber_ab_2018,booth_towards_2013,fraser_finite-size_1996, williamson_elimination_1997, kent_finite-size_1999, lin_twist-averaged_2001, chiesa_finite-size_2006, gaudoin_quantum_2007, drummond_finite-size_2008, kwee_finite-size_2008,dornheim_ab_2016, ceperley_ground_1987, kwon_effects_1998, foulkes_quantum_2001, gurtubay_benchmark_2010, filinov_fermionic_2015, shepherd_many-body_2013, ruggeri_correlation_2018, holzmann_theory_2016, booth_towards_2013, ceperley_ground_1978} 
In all the {\Ia} extrapolations shown in this work, the extrapolation is performed using the following equation: 
\begin{equation}
E(N) = AN^{-1} + E^\mathrm{(1A)}_\mathrm{TDL} %
\label{Scheme_1A_Math}
\end{equation}
where $ E^\mathrm{(1A)}_\mathrm{TDL} $ is the energy at the thermodynamic limit for {\Ia}. 

In {\Ib}, we use an $N^{-2/3}$ convergence rate to extrapolate to the TDL using our basis-set-corrected twist-averaged energies. 
This is also a common extrapolation scheme and is often used for the exchange energy.~\cite{ruggeri_correlation_2018, drummond_finite-size_2008}
For these extrapolations, we used a similar equation to {\Ia}: 
\begin{equation}
E(N) = BN^{-2/3} + E^\mathrm{(1B)}_\mathrm{TDL} %
\label{Scheme_1B_Math}
\end{equation} 
The term $E^\mathrm{(1B)}_\mathrm{TDL}$ is the energy at the thermodynamic limit for this extrapolation scheme. 

In {\IIa}, we use both the $N^{-1}$ and $N^{-2/3}$ convergence rates to extrapolate to the TDL using our basis-set-corrected twist-averaged energies. The equation for the extrapolation, then is a combination of {\Ia} and {\Ib}:
\begin{equation}
E(N) = AN^{-1} + BN^{-2/3} + E^\mathrm{(2A)}_\mathrm{TDL} %
\label{Scheme_2A_Math}
\end{equation}
here, both $A$ and $B$ are free fit parameters that are optimized to give the slopes for the two convergence rates. In theory, the $B$ slope should be similar to the slope of the exchange data, allowing the $BN^{-2/3}$ term, which ultimately cancels the exchange energy convergence in the total energy. 

In {\IIb}, we add a correction term to the $N^{-1}$ extrapolation from {\Ia}. The correction term is derived from the twist-averaged exchange energies, which were collected for a range of system size from $N = 26$ to $826$ for $r_s = 0.1$, $N = 26$ to $508$ for $r_s = 1.0$ and $N = 26$ to $946$ for all other $r_s$. The correction term is derived by fitting the exchange to a $E_\mathrm{x}(N) = -B_\mathrm{x}N^{-2/3} + E_\mathrm{TDL}$ fit. The slope of this fit, $B_\mathrm{x}$, is then incorporated into our $N^{-1}$ extrapolation to the TDL for the correlation energy using the following equation: 
\begin{equation}
E(N) = AN^{-1} + B_\mathrm{x}N^{-2/3} + E^\mathrm{(2B)}_\mathrm{TDL} %
\label{Scheme_2B_Math}
\end{equation}
Here, the $B_\mathrm{x}N^{-2/3}$ term is a correction term to help remove some of the residual FSE in the extrapolation and shifts the TDL energy to be more negative.

{\III} is an equation with three power laws, and is based on the extrapolation scheme presented by Ruggeri \emph{et al}.~\cite{ruggeri_correlation_2018} 

In their paper, they suggested the following relationship: 
\begin{equation}
    E_\mathrm{c} + h_2\N{-2/3} - t_3\N{-1} = c_0 + c_4\N{-4/3} + c_5\N{-5/3} ... %
    \label{Rugerrgi_Math}
\end{equation}
We retained the first three terms of this expansion: %
\begin{equation}
    E(N) = AN^{-4/3} + t_3N^{-1} - h_2N^{-2/3} + E^\mathrm{(3)}_\mathrm{TDL}
\label{Scheme_3_Math}
\end{equation}
Here $t_3$ was the slope taken from Chiesa \emph{et al.} where $t_3 = -\frac{\sqrt{3}}{2}r_s^{-3/2}$ (Ref.~\onlinecite{chiesa_finite-size_2006}) and $h_2$ was given in Drummond \emph{et al.}\cite{drummond_finite-size_2008} as $h_2 = -(\frac{3 C_\mathrm{HF}}{4 \pi r_s})(\frac{1}{4})^{1/3}$, where $C_\mathrm{HF} = 2.837297295$ for the simple cubic UEG.  
The \N{-4/3} term was fit freely with slope $A$.

\subsection{Extrapolation scheme convergence across system size}
\begin{figure}

\subfigure[\mbox{}]{%
\includegraphics[width=0.44\textwidth,height=\textheight,keepaspectratio]{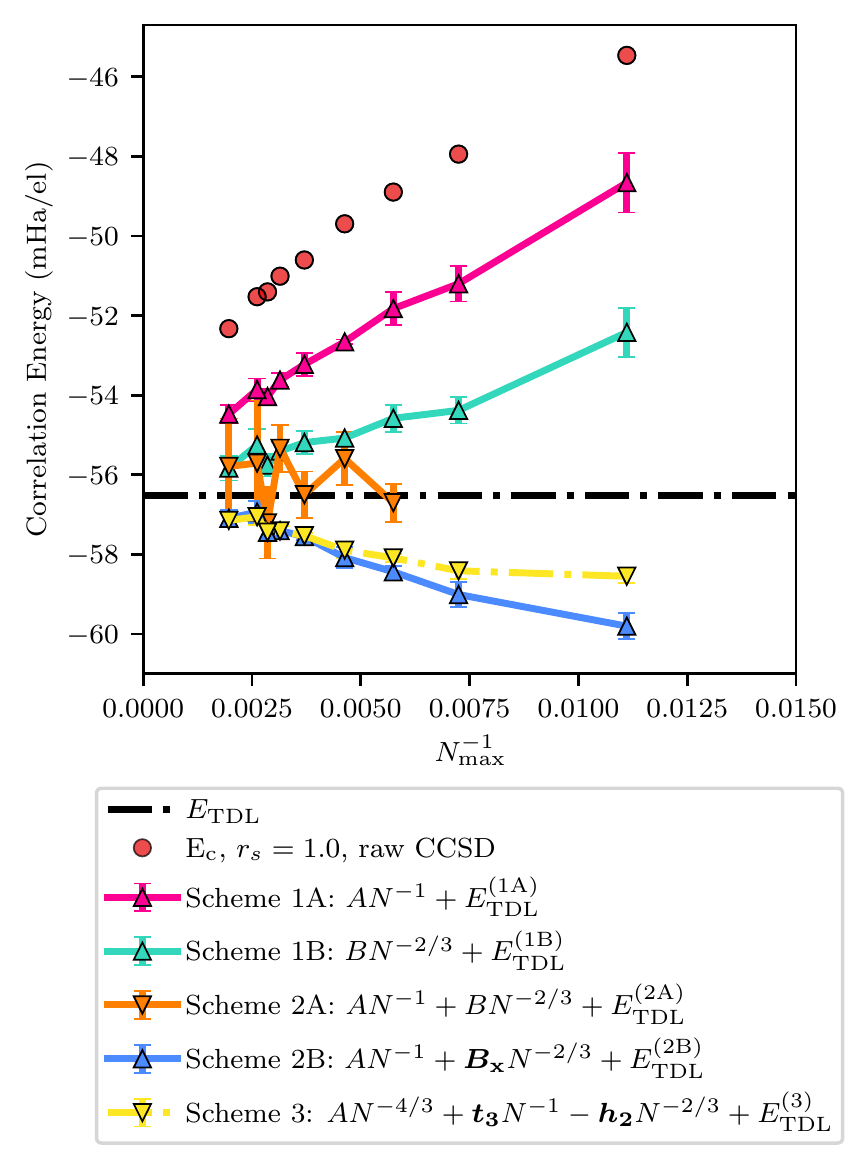}
\label{subfig:WidExp_rs1}
}
\subfigure[\mbox{}]{%
\includegraphics[width=0.45\textwidth,height=\textheight,keepaspectratio]{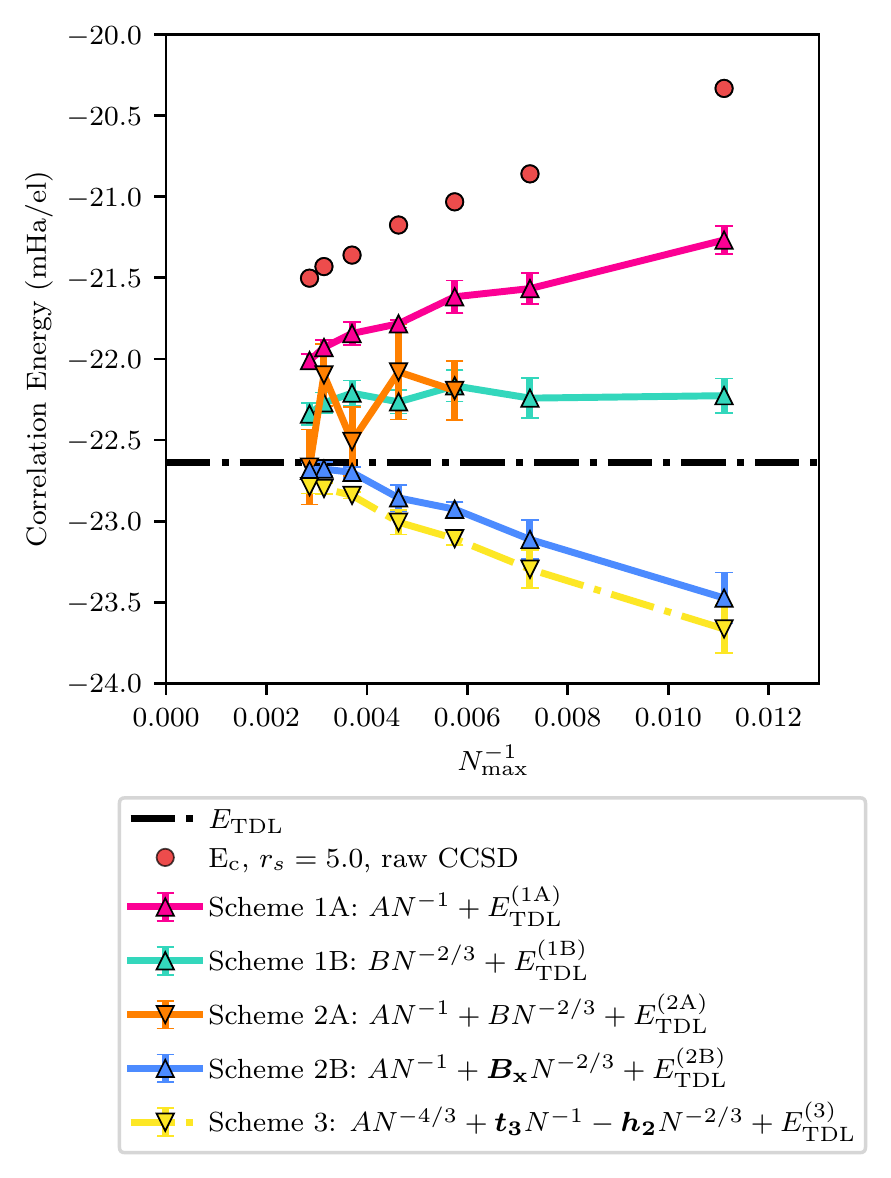}
\label{subfig:WidExp_rs5}
}
\caption{We use a windowed extrapolation technique (see text) to simulate what would happen if we had less data than we actually have. This allows us to evaluate how each of the extrapolation schemes performs.
For each electron number $N_\textrm{max}$, the extrapolation was performed over the $\Delta_i$ largest system sizes $N<N_\textrm{max}$. 
The interval $\Delta_i$ was four for {\Ia}, 1B, 2B, and 3 and six system sizes for {\IIa}.
Each TDL estimate has been graphed again the largest system size (\emph{i.e.} $N_\mathrm{max}$) in the range of system sizes used in the extrapolation.
This is shown for an $r_s$ of (a) $1.0$ and (b) $5.0$.
These are compared to the $E_\textrm{TDL}$ (dash-dotted black line) found by interpolation in \refsec{Calculating_the_Analytical_TDL}. 
In each scheme: $A$ and $B$ are variables found by fitting the correlation energy 
as are all $E_\textrm{TDL}^\textrm{(1A)}$, $E_\textrm{TDL}^\textrm{(1B)}$, \emph{etc.}; $B_X$ is found by fitting the exchange energy; and $h_2$ and $t_3$ (bolded in the figures for emphasis) are found from external sources.~\cite{chiesa_finite-size_2006,drummond_finite-size_2008}
}
\label{fig:Windowed_Extrap}
\end{figure}
Having introduced the extrapolation power laws that we would like to test, 
in this section, we will show the data from our calculations on a variety of UEG systems (\refsec{Calculation_Details}). %
The purpose of this section is to compare the extrapolations schemes shown above with one another. 
In order to compare the effectiveness of extrapolations across different system sizes (modelling an artificial truncation of the data set) we will use a technique we called a windowed extrapolation, where a moving window of points is extrapolated to the TDL using a power law extrapolation scheme. 

These windowed extrapolations were performed as follows.  
Consider $N_i$ to be the $i^\mathrm{th}$ electron number in the data set. 
If $\Delta i$ is the number of points in the window to be extrapolated, then the first and second available extrapolations in the data set are over the interval $i=[1, {1+\Delta i}]$ and $i=[2, {2+\Delta i}]$ respectively. 
The window size, $\Delta i$, is typically chosen as the smallest window size that still offers reasonable errors in the fits. 
The predicted TDL for a single window is assigned to the largest $N$ in the window (i.e. $N_{i+\Delta i}$). %
For our one-variable extrapolations, $\Delta i=4$ was sufficient. For the two-variable extrapolation, $\Delta i=6$ was used instead.
The effectiveness of an extrapolation was then judged by its ability to reproduce the analytically-derived energy value calculated in the previous section (\refsec{Calculating_the_Analytical_TDL}) and the speed of convergence with system size. 
We found that these differences in general varied in size between 0.2 mHa/el and 10 mHa/el.

{%
The results of our windowed extrapolations using the extrapolations schemes from \refsec{Test_Schemes} are shown in \reffig{fig:Windowed_Extrap} for two densities. 
Starting with \reffig{subfig:WidExp_rs1}, where $r_s = 1.0$, we show the convergences of the five extrapolation schemes to the analytical TDL result (from \refsec{Calculating_the_Analytical_TDL}). 
Comparing the different extrapolation schemes, {\Ia} and {\Ib} show the slowest convergence. {{\Ia} does not end up agreeing with the TDL} within the range of electron numbers we studied, {while {\Ib} only agrees at the largest $N$}. 
The other three schemes, {\IIa}, {\IIb} and {\III} have a faster convergence, with {\IIb} and {\III} showing agreement within error to the analytical TDL within the last four points. 
We also see from this graph a reasonable agreement between the predicted TDL values for {\IIb} and {\III} across all the windowed extrapolations, indicating that the three terms in {\III} are accounting for the exchange contribution in {\IIb}. 
{\IIa} shows the quickest convergence to the TDL but has oscillatory convergence due to having a free fit on the {\N{-2/3}} power law.

We see very similar trends with the $r_s = 5.0$ data shown in \reffig{subfig:WidExp_rs5}. 
Here, once again, {\Ia} and {\Ib} have the slowest convergence to the analytical TDL compared with {\IIb} and {\III}. 
In contrast to $r_s = 1.0$ data, however, {\IIa} shows a closer convergence rate to {\Ib} here and a wider spread to the TDL values, which results in a slower convergence to the TDL. 
We also see that the agreement between the {\IIb} and {\III} extrapolated TDL is maintained with this second $r_s$. %
This, again, supports the idea that both schemes are accounting for the \N{-2/3} contribution in the correlation energy.  

}
\begin{table*}[]
\caption{ The differences between the extrapolated thermodynamic limit energy and the analytical (interpolated) thermodynamic limit are shown across schemes for two $r_s$ values.
The TDL energies for each electron number $N_\mathrm{max}$ were obtained using a windowed extrapolation technique (see text). 
For each $N_\mathrm{max}$ shown in the table, the difference between TDL energies was taken such that $\Delta E = E_\mathrm{TDL} - E_\mathrm{Exact}$, where $E_\mathrm{TDL}$ is the predicted thermodynamic limit energies at that $N_\mathrm{max}$ for each extrapolation scheme and $E_\mathrm{Exact}$ is the analytical TDL value at each density. 
For $r_s = 1.0$, the analytical TDL energy is $-56.52$ mHa/el for the correlation energy and $-458.17$ mHa/el for the exchange energy. For $r_s = 5.0$, the analytical TDL energy is $-22.64$ mHa/el for the correlation energy and $-91.63$ mHa/el for the exchange energy. The analytical TDL energies for the correlation and exchange energies were added together at each $r_s$ to get the exchange-correlation analytical TDL value. 
The number in the parenthesis is the error in the difference.   
All energies are in mHa/el.    
}
\resizebox{\columnwidth}{!}{
\renewcommand{\arraystretch}{1.7}
\begin{tabular}{clccccccccc}
      &         & \multicolumn{9}{c}{$N_\mathrm{max}$}          \\ 
\cline{3-11}  
$r_s$ & Scheme & $90$        & $138$      & $174$       & $216$       & $270$       & $318$       & $350$       & $382$      & $508$          \\
\hline
\hline
1.0  & 1A           & 7.9(7)   & 5.3(4)  & 4.7(4)  & 3.86(6) & 3.3(3)   & 2.9(2)  & 2.5(2)  & 2.7(3)   & 2.0(2)   \\
   & 1B           & 4.1(6)   & 2.1(3)  & 1.9(3)  & 1.4(1)  & 1.3(3)   & 1.1(2)  & 0.8(3)  & 1.3(4)   & 0.7(3)   \\
   & 2A          & --       & --      & -0.2(5) & 0.9(7)  & 0.0(6)   & 1.2(6)  & -0.7(9) & 1(2)     & 1(1)     \\
   & 2B          & -3.3(3)  & -2.5(3) & -1.9(1) & -1.6(3) & -1.03(9) & -0.9(2) & -0.9(2) & -0.4(3)  & -0.6(2)  \\
   & 3          & -2.0(2) & -1.9(2) & -1.6(1) & -1.4(2) & -1.01(7) & -0.9(1) & -0.9(2) & -0.5(2) & -0.6(2)   \\
   \hline
   & $E_{x}$, 1A  & -11.3(7) & -8(1)   & -7(1)   & -4.9(9) & -4.8(2)  & -4.1(3) & -3.3(3) & -2.7(3)  & -2.4(2)  \\
   & $E_{x}$, 1B  & -1.1(6)  & -0.6(9) & -0.4(8) & 0.5(6)  & -0.89(8) & -0.7(2) & 0.1(4)  & 0.5(4)   & 0.2(3)   \\
   & $E_{xc}$, 1A & -3.5(1)  & -2.6(6) & -1.9(6) & -1.0(1) & -1.48(7) & -1.2(2) & -0.8(3) & -0.02(2) & -0.35(9) \\
   & $E_{xc}$, 1B & 3(1)     & 1.6(5)  & 1.6(5)  & 2.0(4)  & 0.4(2)   & 0.5(3)  & 0.9(4)  & 1.76(8)  & 0.9(2) \\
\hline
5.0 & 1A           & 1.37(9)  & 1.1(1)  & 1.0(1)   & 0.85(2)   & 0.79(7)  & 0.71(5)  & 0.63(4)  & --  & --  \\
    & 1B           & 0.4(1)   & 0.4(1)  & 0.5(1)   & 0.37(7)   & 0.42(8)  & 0.37(6)  & 0.30(7)   & --  & --  \\
    & 2A          & --       & --      & 0.4(2)   & 0.6(3)    & 0.1(2)   & 0.5(2)   & -0.0(2)  & --  & --  \\
    & 2B          & -0.8(2)  & -0.5(1) & -0.29(4) & -0.22(8)  & -0.06(3) & -0.04(4) & -0.05(5) & --  & --  \\
    & 3          & -1.0(1) & -0.7(1) & -0.47(4) & -0.37(8) & -0.21(2) & -0.16(4) & -0.15(4)  & --  & --  \\
    \hline
    & $E_{x}$, 1A  & -2.3(1)  & -1.6(2) & -1.3(2)  & -0.95(3)  & -0.96(4) & -0.83(6) & -0.65(7) & --  & --  \\
    & $E_{x}$, 1B  & -0.2(1)  & -0.2(2) & -0.1(2)  & 0.2(1)    & -0.18(2) & -0.13(5) & 0.03(7)  & --  & --  \\
    & $E_{xc}$, 1A & -0.91(4) & -0.5(2) & -0.3(1)  & -0.097(8) & -0.16(3) & -0.12(6) & -0.02(8) & --  & --  \\
    & $E_{xc}$, 1B & 0.17(9)  & 0.2(1)  & 0.4(1)   & 0.52(6)   & 0.24(7)  & 0.23(8)  & 0.3(1)   & --  & -- \\
\hline
\end{tabular}
\label{table_windowed_extrap_Diff}
}
\end{table*}

{ 
To make a more detailed comparison, the differences between each of the points for the windowed extrapolations and the analytical TDL values (i.e., $\Delta E = E_\mathrm{TDL} - E_\mathrm{exact}$) are shown in \reftab{table_windowed_extrap_Diff}. %
{%
In this table, we show the results of this difference for all extrapolation schemes at $r_s = 1.0$ and $r_s = 5.0$.
From these results, we see that {\IIb} shows the best comparison to the analytical TDL across both $r_s$. 
Looking at just the $r_s = 1.0$ data, we see that the differences for {\IIb} shown in the table show a fairly steady convergence of the extrapolated TDL to the analytical TDL as $N$ increases, which is the same as what we saw in \reffig{subfig:WidExp_rs1}, with agreement within $1$ mHa/el (within error) reached by $N=270$. %
{\III} for this $r_s$ shows similar trends as {\IIb}. 
In contrast, {\Ia} shows the largest differences to the analytical TDL across all system sizes. %
For {\Ib}, the differences show that there is agreement within $1$ mHa/el (within error) at system sizes as small as $N=270$. %
{\IIa} shows $1$ mHa/el or less agreement to the analytical TDL at the smallest $N$ and maintains this for all $N$, but the differences %
show significant oscillatory behavior as $N$ increases with the largest error seen in the differences across all schemes.

The energy differences from $r_s = 5.0$ show similar results to $r_s = 1.0$, with a few notable differences. 
Here {\IIb} and {\III} are slightly different, with {\IIb} having the smaller differences than {\III} for $N\ge 138$. 
Furthermore, at this density, all of the schemes show differences less than $1$ mHa/el by $N=216$.  
{\IIb} still generally shows the smallest difference out of all five schemes across $N$ starting at $N=174$, with the differences dropping to $<0.1$ mHa/el starting at $N=270$.
{\IIa} can also produce energy difference this low, but shows non-monotonic behavior with large errors as $N$ increases that we were seeing at $r_s = 1.0$, making it a less ideal scheme when extrapolating to the TDL. 
Overall, these results help support the idea that {\IIb} is the best scheme across densities, with {\III} and {\Ib} also working well for smaller densities.   
Interestingly, {\Ia}, which is the most commonly used extrapolation scheme for the correlation energy, performs the worst out of all the schemes, though at low densities (i.e. $r_s= 5$ or greater) this does not seem to matter as much given that all the extrapolation schemes agree within $1$ mHa accuracy. 

}

}

\subsection{One-variable fits of exchange, correlation, and exchange-correlation energies}

\begin{figure}[]
\subfigure[\mbox{}]{%
\includegraphics[width=0.44\textwidth,height=\textheight,keepaspectratio]{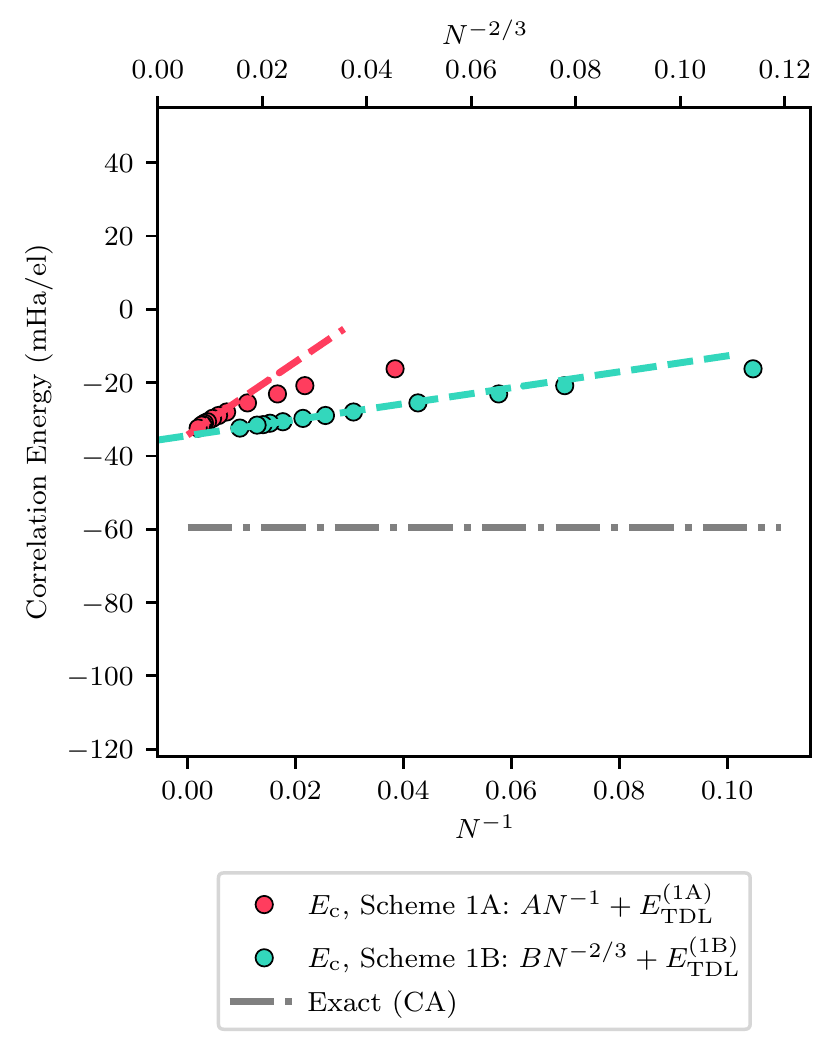}
\label{subfig:Mirror_Cor}
}
\subfigure[\mbox{}]{%
\includegraphics[width=0.45\textwidth,height=\textheight,keepaspectratio]{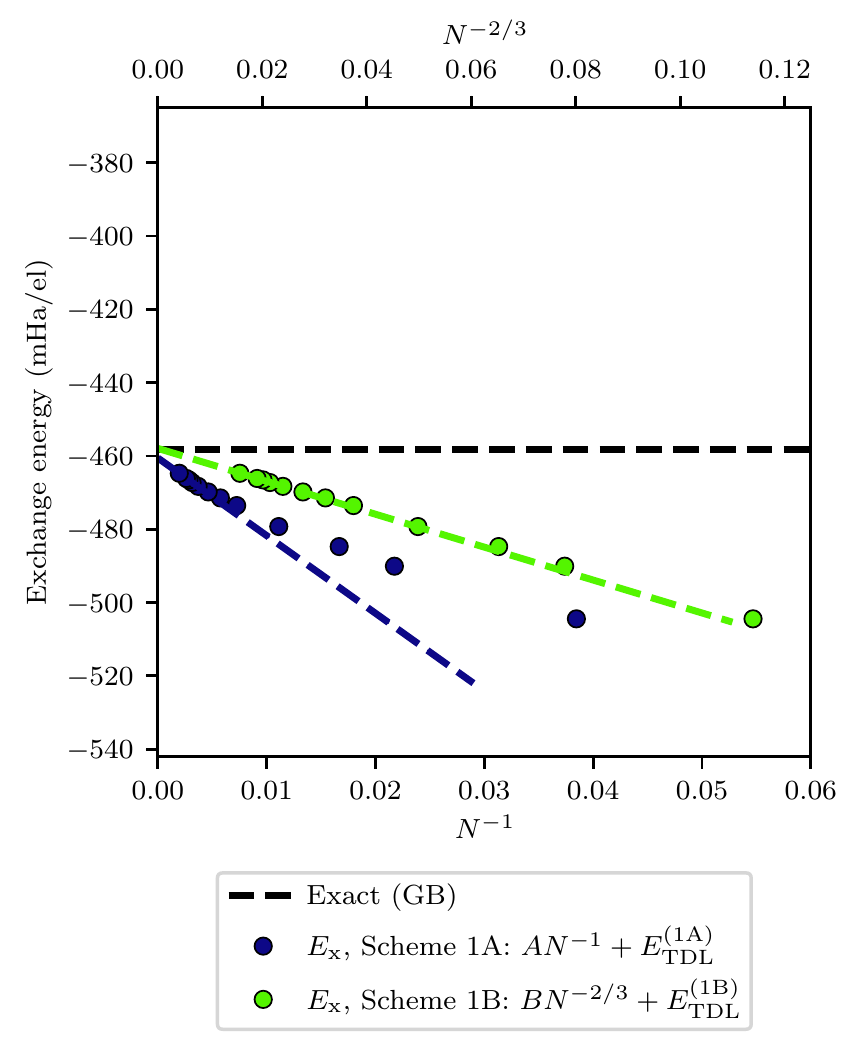}
\label{subfig:Mirror_Ex}
}

\caption{The correlation and exchange energies are shown for a \N{-1} and \N{-2/3} convergence rate for $r_s = 1.0$. 
The exact value for the correlation energy was taken from Ceperley and Alder (CA),~\cite{ceperley_ground_1980, perdew_self-interaction_1981} and is shown as the gray line. 
The exact value for the exchange energy was calculated using the equation provided by Gell-Mann and Brueckner (GB),~\cite{gell-mann_correlation_1957} and is shown as the black line. 
Both the exchange and the correlation energies are shown for a $200$ mHa/el energy range to better show the similarities in the convergences. 
The mirroring in the convergence of the two energies can clearly be seen in both power laws.  
}
\label{fig:Cor-Exchange-Compare_fig}
\end{figure}

There is a tendency for the correlation and exchange energies to mirror one another in how they converge to the TDL. 
This can be most clearly seen by plotting the two as they converge on the same graph (\reffig{fig:Cor-Exchange-Compare_fig}). %
We wanted to investigate the hypothesis that, if exchange and correlation energies were extrapolated over the same range and with the same power laws, the error from using {\Ia} or {\Ib} would cancel. 

Data to investigate this are also shown in  \reftab{table_windowed_extrap_Diff}
as exchange and exchange correlation energies. 
As with the correlation energy, windowed extrapolations were done on both energies with the single-variable power law expansions {\Ia} and {\Ib}; the difference to the analytical TDL was taken. 

In the case of the exchange energy, we calculated the analytical TDL energy using the known result:~\cite{loos_uniform_2016}  %
\begin{equation}
\label{Exchange_exactTDL}
E_\mathrm{x} = -\frac{3}{4\pi} \bigg(\frac{9\pi}{4}\bigg)^{1/3}\frac{1}{r_s}
\end{equation} %
{%
The exact exchange-correlation energy was then calculated as the correlation energy TDL from \refeq{Analytical_TDL} added to the analytical exchange from \refeq{Exchange_exactTDL}.}
{%
Examining the data, we note that the extrapolated exchange TDL underestimates the analytical TDL as convergence is attained. This is in contrast to the fact that the extrapolated correlation TDL systematically overestimates the analytical TDL. 
These trends were observed for both {\Ia} and {\Ib}. 
The sign difference here corresponds to the mirroring seen in 
\reffig{fig:Cor-Exchange-Compare_fig}.

We also investigated directly extrapolating the exchange-correlation energy with the one-variable fits, which was added to \reftab{table_windowed_extrap_Diff}.  
We can see that, overall, this results in {\Ia} generally giving the better TDL estimate (compared to {\Ib}) with the smallest residual error. %
This is consistent with the exchange-correlation energy behaving as \N{-1} into the TDL. 
We can also see from this data that there is a similarity between the result from these extrapolations on the exchange-correlation energy and the result of adding together the residual errors after separate extrapolation of the exchange and correlation energies. 
This suggests that there is a cancellation of error between exchange and correlation energies when they are \emph{both} extrapolated with the \emph{same} schemes, such as a \N{-1} scheme. 
The convergence of {\Ia} on the exchange-correlation energy is consistent with the convergence of {\IIb} on the correlation energy alone.
}

\subsection{Correlation energy TDL across densities}

\begin{figure}[]
\includegraphics[width=0.49\textwidth,height=\textheight,keepaspectratio]{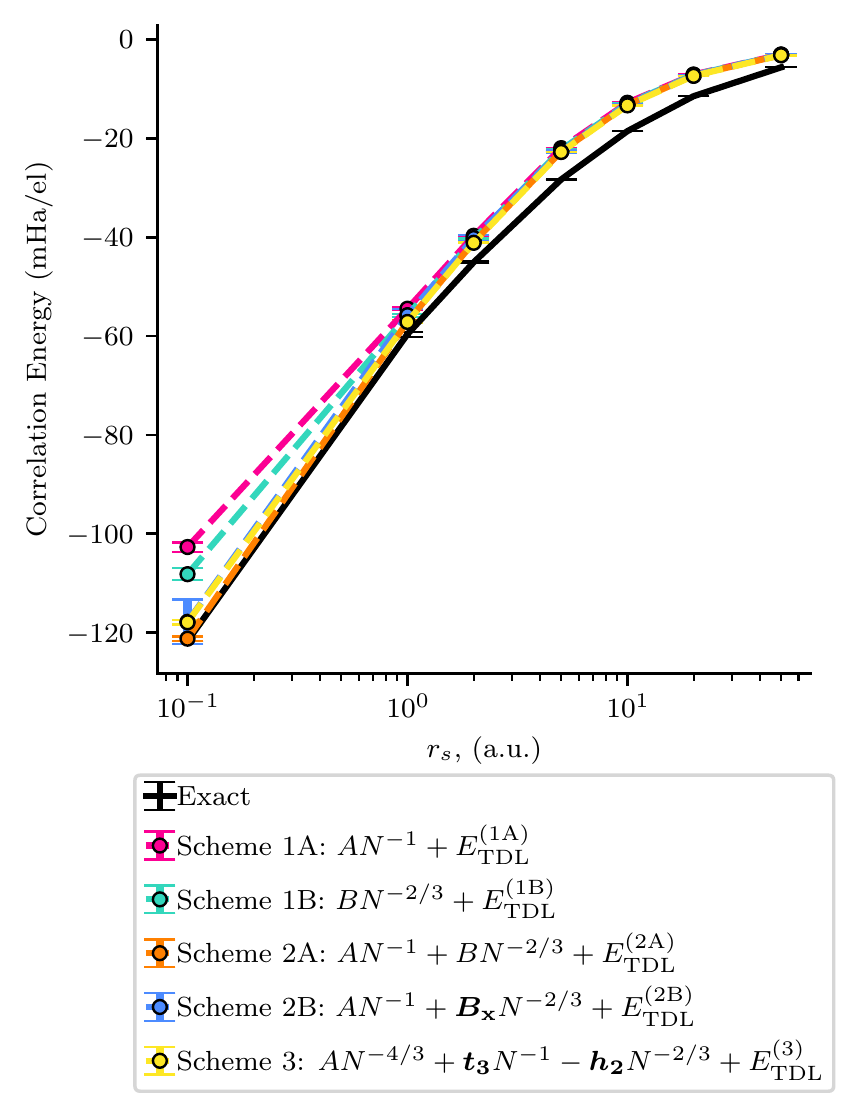}

\caption{The thermodynamic limit correlation energies obtained using our five extrapolation schemes are shown in comparison to the exact correlation energies for a range of densities from $r_s = 0.1$ to $r_s = 50.0$. 
The exact values were calculated from the Ceperley--Alder results ($r_s > 1.0$) and the Gell-Mann--Brueckner results ($r_s < 1.0$) provided by Perdew and Zunger.~\cite{perdew_self-interaction_1981, ceperley_ground_1980, gell-mann_correlation_1957, onsager_integrals_1966} The Ceperley--Alder energies were used to calculate the errors for the exact energies.~\cite{ceperley_ground_1980}
At the higher densities ($r_s = 0.1$ and $1.0$),  
{\IIa}, {\IIb} and {\III} are all shown to reproduce the exact energies. At lower densities ($r_s > 1.0$) the energies for all schemes are shown to differ from the exact energies as is expected for coupled cluster theory.
In each scheme: $A$ and $B$ are variables found by fitting the correlation energy 
as are all $E_\textrm{TDL}^\textrm{(1A)}$, $E_\textrm{TDL}^\textrm{(1B)}$, \emph{etc.}; $B_X$ is found by fitting the exchange energy; and $h_2$ and $t_3$ (bolded in the figures for emphasis) are found from external sources.~\cite{chiesa_finite-size_2006,drummond_finite-size_2008}
}
\label{fig:CorTDL_fig}
\end{figure}
{
In the previous section we showed that, out of the five extrapolation schemes, {\IIb} shows the best comparison to the analytical TDL energies across densities, followed closely by {\III}.
Here, we want to compare the TDL energies from our extrapolation schemes with the exact TDL correlation energies across various densities.   
This comparison will give us more evidence for whether or not {\IIb} is a good general-purpose extrapolation scheme. %
{%
For this comparison, we collected TDL predictions across a range of densities ($r_s = 0.1$ to $50.0$) using an $\fcutM = 2$ for each of our extrapolation schemes described in \refsec{Test_Schemes}. }
Calculation details can be found in \refsec{Calculation_Details}. 
Exact values come from the Ceperley and Alder results ($r_s > 1.0$) and the Gell-Mann and Brueckner results ($r_s < 1.0$) provided by Perdew and Zunger.~\cite{perdew_self-interaction_1981, ceperley_ground_1980, gell-mann_correlation_1957, onsager_integrals_1966} 

}

{
Figure \ref{fig:CorTDL_fig} shows the comparison between the different extrapolation schemes and the exact correlation energies across densities.%
}
All schemes are very similar in their TDL extrapolation at $r_s \ge 5.0$, similar to what we saw in \reftab{table_windowed_extrap_Diff}. 
This is encouraging, as it means that, for sufficiently high electron numbers, the extrapolations all agree at density ranges that are relevant for everyday materials. 
For $r_s < 5.0$, the extrapolation methods begin to have different estimates. At these densities, only {\IIa}, {\IIb} and {\III} are shown to be able to capture the TDL energies within the estimated error from extrapolation.
All three of these, notably, include a contribution from the power law of \N{-2/3}. 
{\IIb} has a lower error from fitting than {\IIa} because it has a fixed coefficient derived from exchange energies. %
{%
}

From these results, it would appear as though {\IIb} is the most consistent in terms of performance when an exchange-energy-slope can be measured. 
We emphasize that this is the slope of the exchange energy after a Madelung term has been added.  Additionally, in our data set, the exchange energy was also twist-averaged, which may influence the quality of the fits. 

\subsection{Practical implications}

From our data and analysis, we make the following suggestions:

\begin{enumerate}

\item Assuming a situation where more exchange energy data is available than correlation energy data, {\IIb} is preferred. This is the scheme where the exchange slope is computed separately, and then included in the correlation energy extrapolation. 
{\IIb} seems especially beneficial at low particle numbers. 

\item If there are comparable amounts of both exchange-energy and correlation-energy data, it is advantageous to extrapolate the exchange-correlation energy directly using {\Ia}. 
Separately extrapolating exchange and correlation energies using a consistent power law (i.e., using {\Ia} or {\Ib} consistently for both exchange and correlation) appears to result in a fortuitous cancellation of error. %
If extrapolations in the literature were to follow the UEG, therefore, this means current extrapolations are likely to be accurate for the total energy. 

\item When the prefactors for both exchange and the leading-order total energy contributions are known, {\III} can also be used to improve the fit quality significantly over the previous power laws at small electron numbers. 

\end{enumerate}

\section{Discussion and concluding remarks}
{%

In conclusion, we incorporated a description of the ground-state (Hartree--Fock) exchange into the transition structure factor of coupled cluster theory. This allowed us to show that there is likely a linear (in $G$) convergence of the transition structure factor $S(G)$ into the origin, rather than the quadratic ($G^2$) convergence described by previous studies. 
Using a new basis set cutoff scheme with our previous twist angle selection approach, we calculated unprecedentedly noiseless energy data into the TDL. This allowed us to investigate and compare five schemes for extrapolating the correlation energy into the TDL.
We find that some accounting for the %
\N{-2/3} term in the extrapolation improves the TDL estimates of the correlation energy.
However, we also showed that if the correlation and exchange energies are both consistently extrapolated with an \N{-1} power law, then the resulting error from using the wrong power law in both cases seems to cancel, at least for the uniform electron gas. 

As this manuscript was under review, it was also noted to us that an analogous observation of $S(G) \sim G$ exists in the literature for the random phase approximation~\cite{bishop_electron_1978}, which they formulate as a ring-diagram-based coupled cluster theory.~%
\cite{scuseria_ground_2008} %
In this work,~\cite{bishop_electron_1978} Bishop and Lurhmann examine the high density electron gas and show analytically that there is a term with appropriate scaling to cancel part of the exchange energy. 
They further make the identification that the energy density in momentum space (\emph{i.e.} components of $\frac{1}{2} S(G) v(G)$ grouped and summed by $G$) scales linearly in $G$. 
Since the number of these terms is proportionate to $G^2$ and $v(G)\propto 1/G^2$, the result is consistent with our observation of $S(G) \sim G$.

We conclude with two limitations of this study. %
First, as with our previous study, we did not employ any finite size corrections prior to using our extrapolation schemes. 
We made this choice as we first wanted to see how the extrapolation schemes behaved without adding in additional corrections when accounting for the %
\N{-2/3} term in the correlation energy.
Second, this study was performed solely on the UEG. It will be important in future studies to show how well these results translate to real materials, including semiconductors and insulators. 
\section{Acknowledgements}

The research presented here was funded by the National Science Foundation under NSF CHE-2045046. 
The University of Iowa is also acknowledged for funding and computer time. 
We thank Michael Mavros for his comments on the manuscript and Andreas Grueneis for showing us Ref.~\onlinecite{bishop_electron_1978}. %
The data set used in this work is available at Iowa Research Online at URL: [to be inserted upon publication].
}

\providecommand{\latin}[1]{#1}
\makeatletter
\providecommand{\doi}
  {\begingroup\let\do\@makeother\dospecials
  \catcode`\{=1 \catcode`\}=2 \doi@aux}
\providecommand{\doi@aux}[1]{\endgroup\texttt{#1}}
\makeatother
\providecommand*\mcitethebibliography{\thebibliography}
\csname @ifundefined\endcsname{endmcitethebibliography}
  {\let\endmcitethebibliography\endthebibliography}{}

 \end{document}